\documentclass[aps,showpacs]{revtex4}

\usepackage{graphics}
\usepackage{graphicx}

\begin{document}

\title{{\it KP}-approach for non-symmetric short-range defects: \\
resonant states and alloy bandstructure}

\author{F.T. Vasko$^*$}
\affiliation{NMRC, University College Cork,
Lee Maltings\\
Prospect Row, Cork, Ireland}

%\date{\today}
%\date{}

\begin{abstract}
The short-range defect with reduced symmetry is studied in the framework
of {\bf kp}-approach taking into account a matrix structure of potential
energy in the equations for envelope functions. The case of the narrow-gap
semiconductor, with defects which are non-symmetric along the $[001]$,
$[110]$, or $[111]$ directions, is considered. Resonant state at a single
defect is analyzed within the Koster-Slater approximation. The bandstructure
modification of the alloy, formed by non-symmetric impurities, is discussed
and a generalized virtual crystal approximation is introduced.
\end{abstract}

\pacs{71.20.Nr; 71.55.Eq}
\maketitle

\section{INTRODUCTION}
Examinations of electron states, localized at short-range defects, began
with the development of the band theory (see Refs. in \cite{1}). The
approach for the calculation of the photoionization cross-section was
formulated in the framework of the effective mass approximation with
short-range impurity potential (Lukovsky model \cite{2}). The short-range
acceptor states were described with the use of the multi-band approach based
on the matrix ${\bf kp}$-Hamiltonian and on the scalar potential energy $U
\delta
({\bf r})$ for the impurity placed at ${\bf r}=0$ \cite{3}. The same
approach was used \cite{4} for the description of both the resonant and
localized electronic states in the gapless and $p$-type materials under
uniaxial stress. The multi-band considerations in \cite{3} and \cite{4} have
demonstrated a good agreement with the experimental data for the
photoionization cross-section of acceptor state and the transformation of
impurity states under uniaxial stress (see discussions in \cite{5} and
\cite{6} respectively). The considerations \cite{2,3,4} were restricted by
the presumption that defect does not mix different bands by itself due to
the scalar character of the impurity potential used. To the best of our
knowledge, a complete description of electronic states at short-range defect,
taking into account {\it a matrix structure} of potential energy, has not
been performed yet. This issue is also of interest in connection with the
last investigations of $A_3B_5$ alloys formed by materials with distinctly
different parameters (see reviews in \cite{7}).

In the present paper, we evaluate the generalized ${\bf kp}$-approach with the
matrix potential energy described a short-range defect with reduced symmetry.
We demonstrate both the appearance of resonant electron state at single defect
and the significant modification of alloy bandstructure due to the non-diagonal
contributions to the matrix potential of defect. The consideration is performed in
the framework of $6\times 6$ Kane model \cite{8}, corresponding to a narrow-gap
semiconductor, for the cases of defect which is non-symmetric along the $[001]$,
$[110]$, or $[111]$ directions. The matrix Green's function of the resonant states
is obtained in the Koster-Slater approximation and the conditions for {\it a
narrow resonant peak} over the conduction or valence band are discussed.
The low-energy electronic states in semiconductor alloy are described by the
Dyson equation with the self-energy function, written in the Born approximation.
Due to the weak damping, we arrive to the effective Hamiltonian with
the non-diagonal matrix of the extreme energies, i.e. we introduce {\it the
generalized virtual crystal approximation}. As a result, the energy spectrum
appears to be anisotropic and splitting of the degenerate hole states takes
place. In addition, a zero-gap case can be realized with an increase of alloy
composition.

The consideration below is organized as follows. In Sec. II we present
the theoretical background including the zero-radius approach for description
of the short-range defect, the Koster-Slater solution for the resonant state
at a single defect, and the generalized virtual crystal approximation for the
low-energy electron states. The results for the resonant states and for the
alloy bandstructure are discussed in Sec. III with the use of the $6\times 6$
Kane model. In Sec. IV we make concluding remarks and discuss the assumptions
made.

\section{BASIC EQUATIONS}
We consider here the equations for the envelope function and for the
single-particle Green's function in the crystal with short-range defects.
Further, we evaluate the contribution of the resonant states into the density
of states and describe the low-energy electronic states using the Dyson
equation with the second-order self-energy function.
\subsection{Zero-radius approach}
The general equation for the envelope wave function, $\Psi_{l{\bf r}}$,
which is introduced through the expansion of an exact wave function,
$\sum_l\Psi_{l{\bf r}}u_{l{\bf r}}$, over the Bloch amplitudes of
$l$th state in the center of Brillouin zone, $u_{l{\bf r}}$, is written
as:
\begin{equation}
\sum_{l'}\left[\hat{H}_{ll'}\Psi_{l'{\bf r}}+\int d{\bf r'}U_{ll'}
({\bf r},{\bf r'})\Psi_{l'{\bf r'}}\right]=E\Psi_{l{\bf r}} .
\end{equation}
Here
$l$ includes band and spin indexes, $\hat{H}_{ll'}$ is the single-electron
{\bf kp}-Hamiltonian described a motion in the ideal crystal. The microscopic
potential of defect, $\delta U_{\bf r}$, appears in the kernel:
\begin{equation}
U_{ll'}({\bf r},{\bf r'})=\int d{\bf r}_1u_{l{\bf r}_1}^*\Delta ({\bf r}
-{\bf r}_1)\delta U_{{\bf r}_1}\Delta ({\bf r}_1-{\bf r'})u_{l'{\bf r}_1} ,
\end{equation}
where $\Delta ({\bf r})=
(1/V)\sum_{\bf k}\exp (i{\bf kr})$ is the 
$\delta$-like function; $V$ is
the normalization volume and $\sum_{\bf k}
\ldots$ is taken over the first Brillouin
zone. Here we consider the short-range 
(localized in the elementary cell)
defect placed at ${\bf r}=0$. For the case 
of the low-energy states, with the
wavelength higher than the lattice constant, 
$a$, we substitute
$\Psi_{l'{\bf r'}=0}$ into the integral term of Eq.(1) and 
this equation
takes form
\begin{equation}
\sum_{l'}\left[\hat{H}_{ll'}\psi_{l'{\bf r}}+U_{ll'}\Delta ({\bf r})
\Psi_{l'{\bf r}=0}\right]=E\Psi_{l{\bf r}} .
\end{equation}
Thus, we have obtained the equations for the envelope wave function with the
matrix potential energy $U_{ll'}\Delta ({\bf r})$ determined through the
components
\begin{equation}
U_{ll'}=\int d{\bf r}u_{l{\bf r}}^*\Delta ({\bf r})\delta U_{\bf r}
u_{l'{\bf r}},
\end{equation}
moreover $U_{ll'}=U_{l'l}^*$. It is necessary to stress that there are no
reasons in order to neglect the non-diagonal components of this matrix if
the local symmetry of $\delta U_{\bf r}$ is reduced in comparison to the
crystal symmetry. This point is a main difference from the standard
{\bf kp}-description of the large-scale potential case and from
the previous consideration of the deep impurities cited above.

Using the potential energy $\sum_j\Delta ({\bf r}-{\bf R}_j)\hat{U}$
for the crystal with $N_{im}$ short-range defects in the volume $V$ placed
at ${\bf R}_j$, $j=1\ldots N_{im}$, we write the equation for the retarded
Green's function in the matrix form
\begin{equation}
\left[ E+i\lambda -\widehat{H}-\sum_j\Delta ({\bf r}-{\bf R}_j)\hat{U}\right]
\hat{G}_{\scriptscriptstyle E}({\bf r},{\bf r'})=\delta ({\bf r}-{\bf r'})
\hat{1},
\end{equation}
where $\lambda\rightarrow +0$ and $\hat{1}$ is the identity matrix. Next, we
transform Eq.(5) into the equation:
\begin{equation}
\hat{G}_{\scriptscriptstyle E}({\bf r},{\bf r'})\simeq
\hat{g}_{\scriptscriptstyle E}({\bf r},{\bf r'})+\sum_j
\hat{g}_{\scriptscriptstyle E}({\bf r},{\bf R}_j)\hat{U}
\hat{G}_{\scriptscriptstyle E}({\bf R}_j,{\bf r'})
\end{equation}
where the free Green's function, $\hat{g}_{\scriptscriptstyle E}({\bf r},
{\bf r'})$, determined from: $(E+i\lambda -\hat{H})
\hat{g}_{\scriptscriptstyle E}({\bf r},{\bf r'})=\delta ({\bf r}-{\bf r'})
\hat{1}$. Thus, the zero-radius approach gives us an infinite (if $N_{im}
\rightarrow \infty$) set of equations for $\hat{G}_{\scriptscriptstyle E}
({\bf R}_j,{\bf r'})$.

\subsection{Koster-Slater approximation}
In order to describe a localized (or resonant) state at $j$th defect, we
rewrite Eq.(6) for $\hat{G}_{\scriptscriptstyle E}({\bf R}_j,{\bf r'})$
neglecting the neighbor defect contributions:
\begin{equation}
\left[ 1-\widetilde{g}_{\scriptscriptstyle E}({\bf R}_j,{\bf R}_j)\hat{U} \right]
\hat{G}_{\scriptscriptstyle E}({\bf R}_j,{\bf r'})\simeq
\hat{g}_{\scriptscriptstyle E}({\bf R}_j,{\bf r'}) .
\end{equation}
The function $\widetilde{g}_{\scriptscriptstyle E}({\bf R}_j,{\bf R}_j)$ means
the free Green's function with the cut-off singularity at coinciding arguments
(i.e. at distances $\sim a$). This approximation is valid if the radius of states is
shorter than $n_{im}^{-1/3}$; $n_{im}=N_{im}/V$ is the concentration of defects.
Substituting $\hat{G}_{\scriptscriptstyle E}({\bf R}_j,{\bf r'})$ determined by
Eq.(7) into Eq.(6) we obtain the Green's function written in the framework of the
Koster-Slater approximation:
\begin{equation}
\hat{G}_{\scriptscriptstyle E}({\bf r},{\bf r'})\simeq
\hat{g}_{\scriptscriptstyle E}({\bf r},{\bf r'})+\sum_j
\hat{g}_{\scriptscriptstyle E}({\bf r},{\bf R}_j)\hat{U}\left( 1-
\hat{\Lambda}_{\scriptscriptstyle E}\hat{U}\right)^{-1}
\hat{g}_{\scriptscriptstyle E}({\bf R}_j,{\bf r'})
\end{equation}
with $\hat{\Lambda}_{\scriptscriptstyle E}\equiv
\widetilde{g}_{\scriptscriptstyle E}({\bf r},{\bf r})$.

The density of states is expressed through $\hat{G}_{\scriptscriptstyle E}
({\bf r},{\bf r'})$ according to the standard formula:
$\rho_{\scriptscriptstyle E}=-Im{\rm tr}\int d{\bf r}
\hat{G}_{\scriptscriptstyle E}({\bf r},{\bf r})/(\pi V)$ where ${\rm tr}\ldots$
means sum over diagonal matrix elements. The impurity contribution to the
density of states, $\Delta\rho_{\scriptscriptstyle E}$, is determined by the
second term of (8). After the permutation of $\hat{g}_{\scriptscriptstyle E}$
under ${\rm tr}\ldots$, we use the equality
$\int d{\bf r}
\hat{g}_{\scriptscriptstyle E}({\bf R}_j,{\bf r})\hat{g}_{\scriptscriptstyle E}
({\bf r},{\bf R}_j)=-[d\widetilde{g}_{\scriptscriptstyle E}({\bf R}_j,{\bf R}_j)
/dE]$ and $\Delta\rho_{\scriptscriptstyle E}$ is transformed into (see similar
formulas in \cite{9,4}):
\begin{equation}
\Delta\rho_{\scriptscriptstyle E}=\frac{n_{im}}{\pi}Im{\rm tr}
\frac{d\hat{\Lambda}_{\scriptscriptstyle E}}{dE}\hat{U}\left(1-
\hat{\Lambda}_{\scriptscriptstyle E}\hat{U}\right)^{-1},
~~~\hat{\Lambda}_{\scriptscriptstyle E}=\frac{1}{V}\sum_{\bf p}{}^{'}
\hat{g}_{\scriptscriptstyle E}({\bf p}) ,
\end{equation}
where $\sum_{\bf p}'\ldots$ means the summation over the region
$|{\bf p}|<\hbar /a$. Here we have also used the free Green's function in
the momentum representation: $\hat{g}_{\scriptscriptstyle E}({\bf p},
{\bf p'})=\delta_{{\bf p},{\bf p'}}\hat{g}_{\scriptscriptstyle E}({\bf p})$
with $\hat{g}_{\scriptscriptstyle E}({\bf p})=( E+i\lambda -
\hat{\varepsilon}-\hat{\bf v}\cdot{\bf p})^{-1}$, where the
${\bf kp}$-Hamiltonian in ${\bf p}$-representation, $\hat{\varepsilon}+
\hat{\bf v}\cdot{\bf p}$, is written through the extreme energy and interband
velocity matrices, $\hat{\varepsilon}$ and $\hat{\bf v}$. It is convenient
to write $\hat{g}_{\scriptscriptstyle E}({\bf p})$ through the dispersion laws
of $k$th band, $\varepsilon_{k{\bf p}}$, according to:
\begin{equation}
\hat{\Lambda}_{\scriptscriptstyle E}=\frac{1}{V}\sum_{k{\bf p}}{}^{'}
\frac{\hat{P}_{k{\bf p}}}{E-\varepsilon_{k{\bf p}}+i\lambda} ,
\end{equation}
where the projection operators onto the $k$th band, $\hat{P}_{k{\bf p}}$,
are given by the matrix elements: $(\hat{P}_{k{\bf p}})_{ll'}=\sum_{\sigma}
\psi_{l{\bf p}}^{(k\sigma )*}\psi_{l'{\bf p}}^{(k\sigma )}$. Here $\sigma$
is the spin index and the column $\psi_{\bf p}^{(k\sigma )}$ is determined
through the eigenstate problem: $(\hat{\varepsilon}+\hat{\bf v}\cdot{\bf p})
\psi_{\bf p}^{(k\sigma )}=\varepsilon_{k{\bf p}}\psi_{\bf p}^{(k\sigma )}$.

\subsection{Generalized virtual crystal approximation}
Consideration of the low-energy electronic states, with the wavelength
exceeds $n_{im}^{-1/3}$, is based on the matrix Dyson equation for the
averaged Green's function. Such equation is obtained from Eq.(6) in the form:
\begin{eqnarray}
\hat{G}_{\scriptscriptstyle E}({\bf p})=\hat{g}_{\scriptscriptstyle E}
({\bf p})+\hat{g}_{\scriptscriptstyle E}({\bf p})
\hat{\Sigma}_{\scriptscriptstyle E}\hat{G}_{\scriptscriptstyle E}({\bf p}), \\
\hat{\Sigma}_{\scriptscriptstyle E}=\frac{n_{im}}{V}\sum_{\bf p}{}^{'}\hat{U}
\hat{G}_{\scriptscriptstyle E}({\bf p})\hat{U}+\ldots , \nonumber
\end{eqnarray}
where the self-energy function, $\hat{\Sigma}_{\scriptscriptstyle E}$, is
written in the self-consistent approximation. Using
$\hat{g}_{\scriptscriptstyle E}({\bf p})$ in
$\hat{\Sigma}_{\scriptscriptstyle E}$
(the Born approximation) we write the
self-energy function through $\hat{\Lambda}_{\scriptscriptstyle E}$ given by
Eq. (10) according to $\hat{\Sigma}_{\scriptscriptstyle E}\simeq n_{im}\hat{U}
\hat{\Lambda}_{\scriptscriptstyle E}\hat{U}$.

Since Eq.(11) is transformed into $\hat{G}_{\scriptscriptstyle E}({\bf p})=
[\hat{g}_{\scriptscriptstyle E}({\bf p})^{-1}-
\hat{\Sigma}_{\scriptscriptstyle E}]^{-1}$, we obtain the averaged density of
states in the form:
\begin{equation}
\overline{\rho}_{\scriptscriptstyle E}=-\frac{Im}{\pi V}\sum_{\bf p}{\rm tr}
\left( E-\hat{\varepsilon}-\hat{\bf v}\cdot{\bf p}-
\hat{\Sigma}_{\scriptscriptstyle E}\right)^{-1}
\end{equation}
and matrix $\hat{\Sigma}_{\scriptscriptstyle E}$ reduces the symmetry at
${\bf p}=0$. According to the consideration below (see Eqs.(16)-(20)),
the self-energy matrix appears to be weakly dependent on $E$ and the damping
contributions to $\hat{\Sigma}$ are also small. Thus, we arrive to the
effective Hamiltonian:
\begin{equation}
\hat{H}^{\scriptscriptstyle (eff)}_{\bf p}=\hat{\varepsilon} +\hat{\Sigma}+
\hat{\bf v}\cdot{\bf p},
~~~\hat{\Sigma}\simeq n_{im}\hat{U}\hat{\Lambda}
\hat{U}
\end{equation}
with the non-digonal matrix $(\hat{\varepsilon}+\hat{\Sigma})$ which determines
the energies of the band extremes. This Hamiltonian corresponds to the
generalized virtual crystal approximation for the alloy formed by low-symmetric
defects: not only the energies of the band extremes are shifted with the alloy
composition (see \cite{10}; note, that $\hat{\Sigma}\propto n_{im}$ in the Born
approximation), but also the symmetry of $\hat{H}^{\scriptscriptstyle (eff)}
_{\bf p}$ is reduced
due to the non-diagonal contributions from $\hat{\Sigma}$.
The characteristic
equation correspondent to the effective Hamiltonian (13) is 
written in the form:
\begin{equation}
{\rm det}|\hat{\varepsilon} +\hat{\Sigma}+\hat{\bf v}\cdot{\bf p}-E|=0
\end{equation}
and the dispersion laws appear to be anisotropic, with splitted valence bands,
like for the case of stressed semiconductors.

\section{Results}
Let us turn to Eqs. (9) and (14) for the narrow-gap semiconductor described
in the framework of the isotropic $6\times 6$ Kane model. For the sake of
simplicity, we restrict our consideration to the model of defect with the  
microscopic potential, $\delta U_{\bf r}$, which is only non-symmetric along 
the $[001]$ (case $A$), $[110]$ (case $B$), or $[111]$ (case $C$) directions.

\subsection{$6\times 6$ Kane model}
The matrices $\hat{U}$ and $\hat{\Lambda}_{\scriptscriptstyle E}$ are determined
below using the set of Bloch functions which are expressed according
to 
\cite{11} in terms of the
periodic basis $|S\rangle$, $|X\rangle$, $|Y\rangle$, 
and $|Z\rangle$. Within the above approximations, the potential matrix (4) is
written in the form:
\begin{eqnarray}
\hat{U}=\left|\begin{array}{ll} \hat{u}_c & \widetilde{u} \\ \widetilde{u}^+ &
\hat{u}_v \end{array}\right| ,~~~~ \widetilde{u}_{\scriptscriptstyle A}=u_z
\left|\begin{array}{llll} ~~0 & -i\sqrt{\frac{2}{3}} & 0 & 0 \\ i\sqrt{\frac{2}
{3}} & ~~~~0 & 0 & 0 \end{array}\right|, \nonumber \\
\widetilde{u}_{\scriptscriptstyle B}=u_d\left|\begin{array}{llll} \frac{-i}
{\sqrt{3}}e_+ & ~0 & ie_- & ~~0 \\ ~0 & \frac{-i}{\sqrt{3}}e_- & ~0 & -ie_+
\end{array}\right| , \\
\widetilde{u}_{\scriptscriptstyle C}=u_o\left|\begin{array}{llll} \frac{-i}
{\sqrt{3}}e_+ & -i\sqrt{\frac{2}{3}} & ie_- & ~0 \\ i\sqrt{\frac{2}{3}} &
~\frac{-i}{\sqrt{3}}e_- & ~0 & ie_+ \end{array}\right|  .\nonumber
\end{eqnarray}
Here $\hat{u}_c$ and $\hat{u}_v$ are proportional to the $2\times 2$ and
$4\times 4$ identity matrices: $\hat{u}_c=\hat{1}\langle S|\delta U|S\rangle
\equiv\hat{1}u_c$ and $\hat{u}_v=\hat{1}\langle X|\delta U|X\rangle\equiv\hat{1}
u_v$; the non-diagonal part of $\hat{U}$ is expressed through the $4\times 
2$
matrix, $\widetilde{u}$, where $e_{\pm}=
(1\pm i)/\sqrt{2}$. The matricis 
$\widetilde{u}_{\scriptscriptstyle A,B,C}$ are proportional to $u_z\equiv
\langle S|\delta U|Z\rangle\neq 0$ with $\langle S|\delta U|X\rangle =\langle S|
\delta U|Y\rangle
=0$ (case $A$), or $u_d\equiv\langle S|\delta U|X\rangle =
\langle S|\delta U|Y\rangle\neq 0$ with $\langle S|\delta U|Z\rangle =0$ (case 
$B$), or $u_o\equiv
\langle S|\delta U|X\rangle =\langle S|\delta U|Y\rangle =
\langle S|\delta U|Z\rangle\neq 0$ (case $C$).

Since the dispersion laws $\varepsilon_{kp}$ are isotropic, we use in Eq.(10)
the projection operators which are averaged over the angle, so that
$\hat{P}_{k{\bf p}}$ are proportional to $\delta_{ll'}$. Thus, the diagonal
matrix $\hat{\Lambda}_{\scriptscriptstyle E}$ is determined by the matrix
elements
\begin{equation}
(\hat{\Lambda}_{\scriptscriptstyle E})_{ll}=\left\{\begin{array}{ll}
\lambda_{c{\scriptscriptstyle E}}-i\eta_{c{\scriptscriptstyle E}}, & l=1,2 \\
\lambda_{v{\scriptscriptstyle E}}-i\eta_{v{\scriptscriptstyle E}}, & l=3-6
\end{array}\right. ,
\end{equation}
where $\lambda_{k{\scriptscriptstyle E}}$ are weakly dependent on $E$ due to
the dominant contribution from $|{\bf p}|\sim\hbar /a$. Because of this, we use
the expansion $\lambda_{k{\scriptscriptstyle E}}\simeq \lambda_k-l_k(E+
\varepsilon_g/2)$ where the energy $-\varepsilon_g/2$ corresponds the middle of
gap and the coefficients $\lambda_k$ and $l_k$ are given by
\begin{eqnarray}
\lambda_c=-\frac{1}{V}\sum_{q{\bf p}}{}^{'}\frac{|C_{qp}|^2}{\varepsilon_g/2+
\varepsilon_{qp}} , ~~~l_c=\frac{1}{V}\sum_{q{\bf p}}\frac{|C_{qp}|^2}
{(\varepsilon_g/2+\varepsilon_{qp})^2} , \nonumber \\
\lambda_v=-\frac{1}{V}\sum_{q{\bf p}}{}^{'}\frac{|C_{qp}|^2a_{qp}}{\varepsilon_g/2+
\varepsilon_{qp}}-\frac{1}{V}\sum_{\bf p}{}^{'}(\varepsilon_g/2+\varepsilon_{hp})
^{-1} , \\
l_v=\frac{1}{V}\sum_{q{\bf p}}\frac{|C_{qp}|^2a_{qp}}{(\varepsilon_g/2+
\varepsilon_{qp})^2}+\frac{1}{V}\sum_{\bf p}(\varepsilon_g/2+\varepsilon_{hp})
^{-2} . \nonumber
\end{eqnarray}
Here $\sum_q\ldots$ means summation over the electron ($c$-) and light-hole
($lh$-) state while the heavy hole ($hh$-) state, with the dispersion law
$\varepsilon_{hp}$, appears in $\lambda_v$ and $l_v$. Note also, that the
coefficients $l_k$ have no divergence at $|{\bf p}|\rightarrow\infty$. Eqs.
(17) are written through the coefficient, $a_{qp}$, and the normalization
factor, $C_{qp}$, which are  determined as follows:
\begin{equation}
a_{qp}=\frac{({\cal P}p)^2}{3(\varepsilon_{qp}+\varepsilon_g)^2}, ~~~|C_{qp}|^2
=\left[1+\frac{2({\cal P}p)^2/3}{(\varepsilon_{qp}+\varepsilon_g)^2}
\right]^{-1},
\end{equation}
where $\cal P$ is the Kane velocity. The broadening factors in Eq.(16),
$\eta_{k{\scriptscriptstyle E}}$, are expressed through the $k$th band density
of states, $\rho_k(E)=(2/V)\sum_{\bf p}\delta (E-\varepsilon_{kp})$, according
to
\begin{equation}
\eta_{c{\scriptscriptstyle E}}=\frac{\pi}{2}\sum_q|C_{qp}|^2_{\varepsilon_{qp}
={\scriptscriptstyle E}}\rho_q(E), ~~~\eta_{v{\scriptscriptstyle E}}=\frac{\pi}
{2}\sum_q(|C_{qp}|^2a_{qp})_{\varepsilon_{qp}={\scriptscriptstyle E}}\rho_q(E)
+\frac{\pi}{2}\rho_{hh}(E) .
\end{equation}
Thus, the model under consideration involves five phenomenological parameters: 
the potentials $u_{c,v}$ and $u_z$ (case A), or $u_d$ (case B), or $u_o$ 
(case C) in Eq.(15) and the factors $\lambda_{c,v}$ determined by Eq.(17).
Another values ($l_{c,v}$ in Eq.(17) and the functions $\eta_{c,v
{\scriptscriptstyle E}}$ given by Eq. (19)) are expressed through the
parameters of the ideal crystal determined in $\bf kp$-approach.

\subsection{Resonant states}
We consider here $\Delta\rho_{\scriptscriptstyle E}$ determined by Eq.(9) taking
into account both diagonal and non-diagonal contributions to Eq.(15). Below we
use $6\times 6$ matrix $\hat{\Lambda}_{\scriptscriptstyle E}\hat{U}$ and we
introduce $(1-\hat{\Lambda}_{\scriptscriptstyle E}\hat{U})^{-1}$ in the
matrix form:
\begin{equation}
\hat{\Lambda}_{\scriptscriptstyle E}\hat{U}=\left|\begin{array}{ll}
\lambda_{c{\scriptscriptstyle E}}u_c & \lambda_{c{\scriptscriptstyle E}}
\widetilde{u} \\ \lambda_{v{\scriptscriptstyle E}}\widetilde{u}^+ &
\lambda_{v{\scriptscriptstyle E}}u_v \end{array}\right| ,~~~~ (1-
\hat{\Lambda}_{\scriptscriptstyle E}\hat{U})^{-1}\equiv\left|\begin{array}{ll}
\hat{x}_c & \hat{x}_{cv} \\ \hat{x}_{vc} & \hat{x}_v \end{array}\right|
,
\end{equation}
where the $4\times 2$ matrices $\widetilde{u}_{\scriptscriptstyle A,B,C}$ are
given by the Eq.(15). The components of the reciprocal matrix in (20) are
determined from the $6\times 6$ matrix equation:
\begin{equation}
\left( 1-\left|\begin{array}{ll} \lambda_{c{\scriptscriptstyle E}}u_c &
\lambda_{c{\scriptscriptstyle E}}\widetilde{u} \\
\lambda_{v{\scriptscriptstyle E}}\widetilde{u}^+ &
\lambda_{v{\scriptscriptstyle E}}u_v \end{array}\right|\right)\left|
\begin{array}{ll} \hat{x}_c & \hat{x}_{cv} \\ \hat{x}_{vc} & \hat{x}_v
\end{array}\right| =\hat{1} .
\end{equation}

Using these notations, we write Eq.(9) as follows:
\begin{equation}
\Delta\rho_{\scriptscriptstyle E}=\frac{n_{im}}{\pi}Im\left[
\lambda_{c{\scriptscriptstyle E}}'{\rm tr}\left( u_c\hat{x}_c+\widetilde{u}
\hat{x}_{vc}\right) +\lambda_{v{\scriptscriptstyle E}}'{\rm tr}\left( u_v
\hat{x}_v+\widetilde{u}^+\hat{x}_{cv}\right)\right] .
\end{equation}
where $\lambda_{q{\scriptscriptstyle E}}'\equiv d\lambda_{q
{\scriptscriptstyle E}}/dE$. Further calculations reduce Eq.(21) to the $2
\times 2$ linear equations and result for $\Delta\rho_{\scriptscriptstyle E}$
takes standard form \cite{4,8}:
\begin{equation}
\Delta\rho_{\scriptscriptstyle E}=\frac{2n_{im}}{\pi}Im\left[\frac{(d{\cal L}
_{\scriptscriptstyle E}/dE)}{{\cal L}_{\scriptscriptstyle E}}+\frac
{\Lambda_{v{\scriptscriptstyle E}}'u_v}{1-\Lambda_{v{\scriptscriptstyle E}}u_v}
\right]
,
\end{equation}
where the last term is due to the heavy hole contribution. The function
${\cal L}_{\scriptscriptstyle E}$ is written as
\begin{equation}
{\cal L}_{\scriptscriptstyle E}=(1-\lambda_{c
{\scriptscriptstyle E}}u_c)(1-\lambda_{v{\scriptscriptstyle E}}u_v)-w
\lambda_{c{\scriptscriptstyle E}}\lambda_{v{\scriptscriptstyle E}},
~~~~w=\left\{\begin{array}{ll} 2u_z^2/3, & A \\ 4u_d^2/3, & B \\
2u_o^2, & C \end{array} \right.
\end{equation}
with the different factors $w$ for the cases $A$-$C$ under consideration.

According to Eqs.(16)-(19), ${\cal L}_{\scriptscriptstyle E}$ depends on $E$
weakly and has the small imaginary part. It is convenient to
introduce the level energy $E_o$ which is given by a root of the linear
equation: $Re{\cal L}_{\scriptscriptstyle E_o}=0$. Thus, the function
${\cal L}_{\scriptscriptstyle E}$ is transformed into ${\cal L}_0'(E-E_o-i
\Gamma_o)$ where the coefficient, ${\cal L}_0'$, and the broadening energy,
$\Gamma_o$, are introduced as follows:
\begin{eqnarray}
{\cal L}_0'=l_cu_c(1-\lambda_vu_v)+l_vu_v(1-\lambda_cu_c)+w(l_c\lambda_v+l_v
\lambda_c),  \nonumber \\
\Gamma_o=[\eta_{c{\scriptscriptstyle E_o}}u_c+\eta_{v{\scriptscriptstyle E_o}}
u_v+w(\eta_{c{\scriptscriptstyle E_o}}\lambda_v+\eta_{v{\scriptscriptstyle E_o}}
\lambda_c)]/{\cal L}_0' .
\end{eqnarray}
Note that $\eta_{k{\scriptscriptstyle E}}$ and $\Gamma_o$ are proportional to 
the density of states in $c$- or $v$-band, if
$E_o>0$ or $E_o<-\varepsilon_g$. 
It follows that the broadening energy for the $v$-band resonant state,
with 
$E_o<-\varepsilon_g$, exceeds $\Gamma_o$ for the $c$-band resonant state,
with 
$E_o>0$.
The same transformation can be performed for the last term of Eq.(23), 
so
that the level, $E_v$, is determined from the equation 
$1-
\lambda_{v{\scriptscriptstyle E}}u_v=0$ and the broadening energy is equal
to $\eta_{v{\scriptscriptstyle E}_v}/l_v$.

If a deep local level appears in the gap, $0>E_{o,v}>-\varepsilon_g$, then
the broadening energy is replaced by $+0$ and the contribution to
$\Delta\rho_{\scriptscriptstyle E}$ takes form: $2n_{im}\delta (E-E_{o,v})$.
The cases $E_{o,v}>0$ or $E_{o,v}<-\varepsilon_g$ are corresponded to a
resonant state over $c$- or $v$-band and $\Delta\rho_{\scriptscriptstyle E}$
has the Lorentzian shape with the broadening energy $\Gamma_o$:
\begin{equation}
\Delta\rho_{\scriptscriptstyle E}=\frac{2n_{im}}{\pi}\frac{\Gamma_o}
{(E-E_o)^2+\Gamma_o^2} .
\end{equation}
The same result with the level energy $E_v$ and the broadening energy
$\eta_{v{\scriptscriptstyle E}_v}/l_v$ is valid for the $hh$-contribution.
Here and in Eq.(24) we have supposed that
$E_{o,v}$ or $|E_{o,v}-\varepsilon_g|$ exceed the broadening energies for
the resonant states over $c$- or $v$-bands respectively. Both $E_o$ and
$\Gamma_o$ in Eq.(26) are determined by the short-range contributions to the
factors $\lambda_{c,v}$, so that neither an absolute value of broadening
energy nor a sign of $\Gamma_o$, which determines peak- or dip-modification
in the density of states, are not fixed in the consideration performed. The
only dependence on $\varepsilon_g $ (i.e. on hydrostatic pressure) can be
evaluated from these results within a few fitting parameters.

\subsection{Alloy bandstructure}

Next, we examine modifications of the alloy bandstructure due to contributions
into Eq.(14) from the matrix self-energy function determined by the Eqs.(15),
(20). Using $2\times 2$ matrix notations, we write $\hat{\varepsilon}+
\hat{\Sigma}$ in the form:
\begin{equation}
\left|\begin{array}{lll} \hat{E}_1 & 0 & 0 \\ 0 & \hat{E}_3 & \hat{\Delta} \\
0 & \hat{\Delta}^+ & \hat{E}_5 \end{array}\right|
,
\end{equation}
where the non-diagonal component, $\hat{\Delta}$, is determined for the cases
$A$-$C$ through the characteristic energies $\delta =n_{im}u_d^2\lambda_c/
\sqrt{3}$ and $\bar{\delta} =n_{im}u_o^2\lambda_c/\sqrt{3}$ according to:
\begin{equation}
\hat{\Delta}_{\scriptscriptstyle A}=0, ~~~\hat{\Delta}_{\scriptscriptstyle B}=
i\delta , ~~~\hat{\Delta}_{\scriptscriptstyle C}=\bar{\delta}\left|\begin{array}
{ll} ~~~~i & -(1+i) \\ -(1-i) & ~~~~i \end{array}\right| .
\end{equation}
The diagonal matrices $\hat{E}_{1,3,5}$ are determined by the components
presented in the Table.
\begin{table}
\begin{tabular}{|c|c|c|c|}
\hline
& $E_{1,2}$ & $E_{3,4}$ & $E_{5,6}$ \\ \hline
$A$ & $n_{im}(u_c^2\lambda_c+u_z^2\lambda_v)$ & $-\varepsilon_g+n_{im}(u_v^2
\lambda_v+u_z^2\lambda_c)$ & $-\varepsilon_g+n_{im}u_v^2\lambda_v$ \\ \hline
$B$ & $n_{im}(u_c^2\lambda_c+4u_d^2\lambda_v/3)$ & $-\varepsilon_g+
n_{im}(u_v^2\lambda_v+u_d^2\lambda_c/3)$ & $-\varepsilon_g+n_{im}(u_v^2
\lambda_v+u_d^2\lambda_c)$ \\ \hline
$C$ & $n_{im}(u_c^2\lambda_c+2u_o^2\lambda_v)$ & $-\varepsilon_g+n_{im}
(u_v^2\lambda_v+u_o^2\lambda_c)$ & $-\varepsilon_g+n_{im}(u_v^2\lambda_v+u_o^2
\lambda_c)$ \\ \hline
\end{tabular}
\caption{\label{tab:table1}Components of diagonal matrices $\hat{E}_{1,3,5}$ in
 Eq.(27).}
\end{table}

 \begin{figure}
\begin{center}
\includegraphics{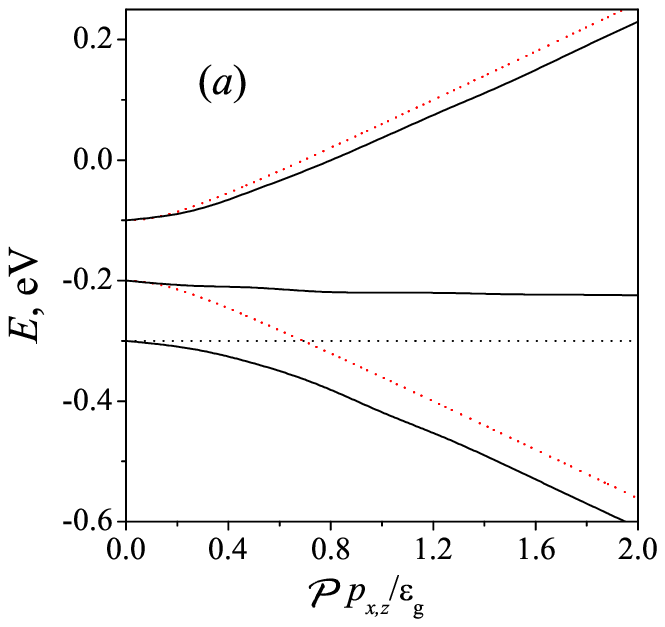}
\includegraphics{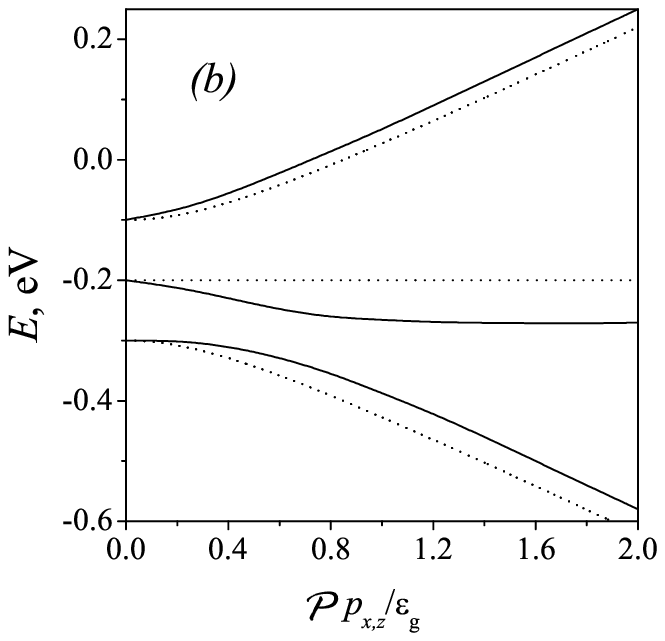}
\end{center}
\addvspace{-1 cm}\caption{The dispersion laws along ${\bf p}\bot OZ$ ($p_x$,
solid curves) and ${\bf p}\| OZ$ ($p_z$, dotted curves) directions for the case $A$,
if $E_3>E_5$ ($a$) and $E_3<E_5$ ($b$).}
\label{fig.1}
\end{figure}

\begin{figure}
\begin{center}
\includegraphics{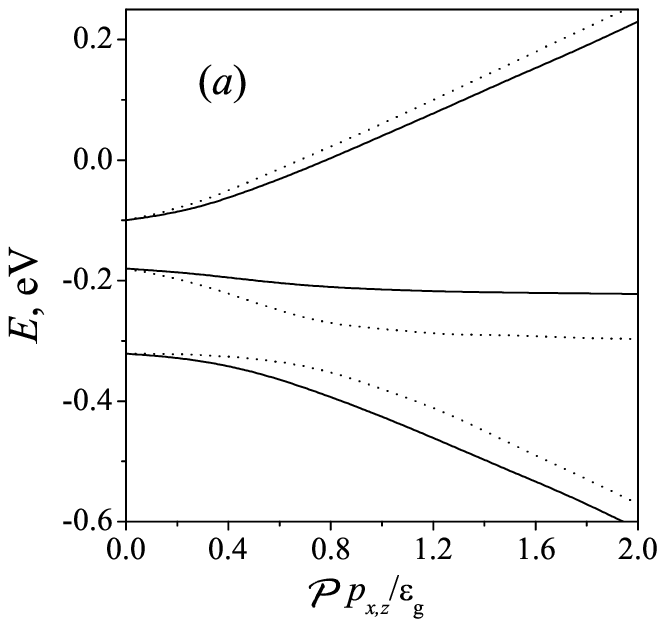}
\includegraphics{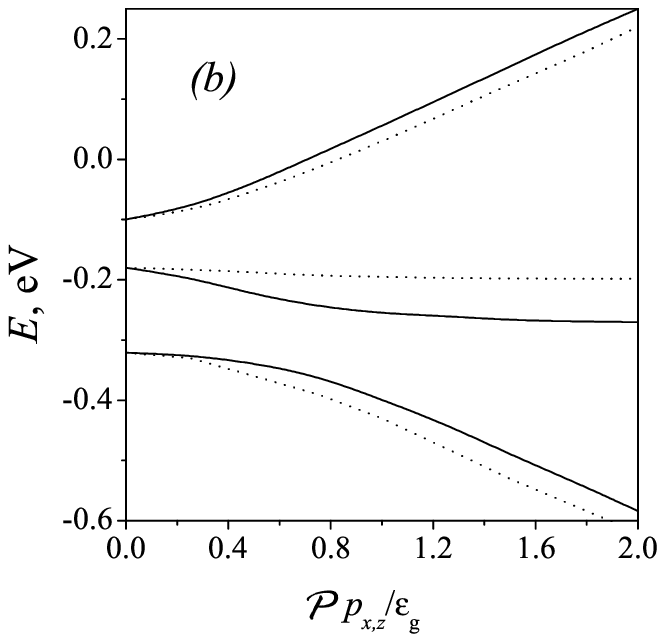}
\end{center}
\addvspace{-1 cm}\caption{The same as in Fig.1 for the case $B$.}
\label{fig.2}
\end{figure}

The spin-degenerate dispersion equations are obtained from the Eqs.(14), (27)
and (28) as follows:
\begin{eqnarray}
\frac{({\cal P}p_{\scriptscriptstyle \bot})^2}{2}(E-E_3)+\left[\frac{({\cal P}
p_{\scriptscriptstyle \bot})^2}{6}+
\frac{2}{3}({\cal P}p_z)^2\right] (E-E_5)=
(E-E_1)\times\left\{\begin{array}{ll}(E-E_3)(E-E_5), & (A) \\ (E-E_3)(E-E_5)-
\delta^2, & (B) \end{array}
\right. \nonumber \\
\frac{2}{3}({\cal P}p)^2(E-E_3)+\frac{2}{3}\bar{\delta}{\cal P}^2(p_xp_y
+p_xp_z+p_yp_z)=(E-E_1)\left[ (E-E_3)^2-\bar{\delta}^2\right] , ~~~~(C) ,
\end{eqnarray}
where $p_{\scriptscriptstyle \bot}^2=p_x^2+p_y^2$. According to Eq.(29) the 
extreme energies,
$E_{3,5}$ in the case $A$ and $E_{\scriptscriptstyle \pm}=E_3
\pm |\delta |$ in the case $B$,
appear to be linear dependent on the alloy 
composition \cite{12}. A bowing
of $v$-band extreme energies with increase in 
alloy composition takes place
for the case $B$
\begin{equation}
E_{\scriptscriptstyle \pm}=\frac{E_3+E_5}{2}\pm\sqrt{\left(\frac{E_3+E_5}{2}
\right)^2+
\delta^2}
\end{equation}
due to the non-diagonal contributions in Eq.(27). The anisotropy of energy
spectrum takes place if $E_3\neq E_5$ in the cases $A$ and $B$, or if
$\bar{\delta}\neq 0$ in the case $C$. These dispersion laws are shown in
Figs.1-3 for the narrow-gap alloy with $2m_o{\cal P}^2=$25 eV, $m_o$ is the
free electron mass, and with $\varepsilon_g=$0.25 eV. Here we suppose the
negative shift of $c$-extremum, $E_1=-$0.1 eV, for the all cases $A$-$C$
and the energies $\delta$ , $\bar{\delta}$ are chosen as 0.05 eV. We
consider a different order of $v$-bands: $E_3=-0.2$ eV, $E_5=-0.3$ eV $(a)$
and $E_3=-0.3$ eV, $E_5=-0.2$ eV  $(b)$ for the cases $A$, $B$ in Figs.1,2,
and we suppose $E_{3-6}=-0.3$ eV and $E_{3-6}=-0.2$ eV in Figs. 3$a$ and 3$b$
respectively. One can see that the dispersion laws appear to be weakly
anisotropic in the cases $A$ and $B$ while a visible anysotropy takes place
for the case $C$. We note also a substantial modification of the $v$-band
extrema if the splitting energy, $|E_3-E_5|$, is comparable with
$\varepsilon_g$.

\begin{figure}
\begin{center}
\includegraphics{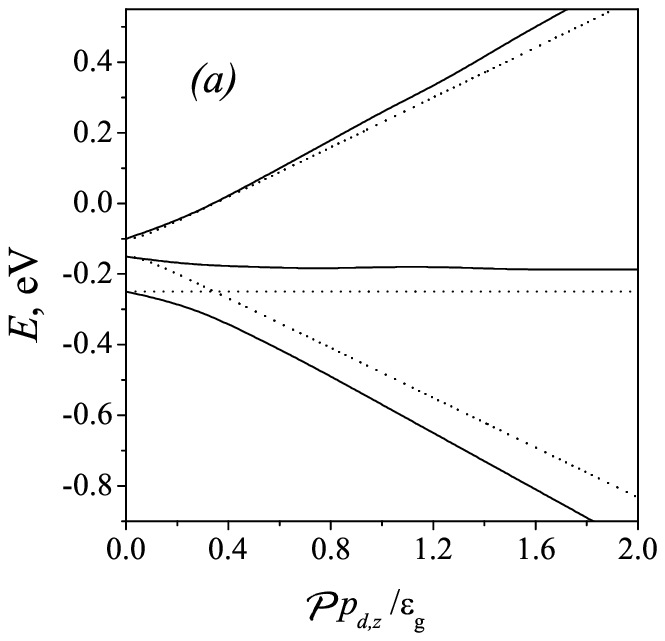}
\includegraphics{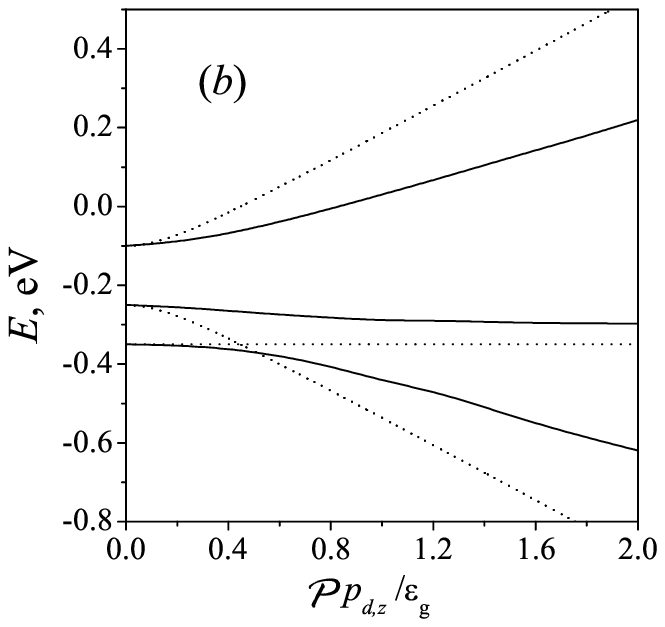}
\end{center}
\addvspace{-1 cm}\caption{The dispersion laws along the cubic diagonal
${\bf p}\| [111]$ ($p_d$, solid curves) and ${\bf p}\| OZ$ ($p_z$, dotted
curves) for the cases $C$ if $E_{3,5}=-0.2$ eV ($a$) and $E_{3,5}=-0.3$ eV
($b$).}
\label{fig.3}
\end{figure}

The zero-gap bandstructure can be realized if $E_1=E_3$ or if $E_1=E_5$ for
the case $A$ and if $E_1=E_+$ for the case $B$ and if $E_1=E_3+\bar{\delta}$
for the case $C$. In the vicinity of the cross-point energy, if $|E-E_1|,
{\cal P}|{\bf p}|$ are less than the energy spacing to the lower $v$-band,
the dispersion laws are given by:
\begin{eqnarray}
E_{{\scriptscriptstyle \pm}{\bf p}}\simeq E_1\pm\left\{\begin{array}{ll}
\sqrt{({\cal P}p_{\scriptscriptstyle \bot})^2/6+(2/3)({\cal P}p_z)^2} , & 
E_1=E_3 \\ \pm {\cal P}p_{\scriptscriptstyle \bot}/\sqrt{2}, &
E_1=E_5 \end{array}\right. ~~~(A) \nonumber \\
E_{{\scriptscriptstyle \pm}{\bf p}}\simeq E_1\pm{\cal P}\sqrt{\frac{E_1-E_5}
{E_1-E_-}\left(
\frac{p_{\scriptscriptstyle \bot}^2}{6}+\frac{2p_z^2}{3} 
\right) +\frac{E_1-E_3}{E_1-E_-}
\frac{p_{\scriptscriptstyle \bot}^2}{2}}, 
~~~(B)\\
E_{{\scriptscriptstyle \pm}{\bf p}}\simeq E_1\pm\frac{\cal P}{\sqrt{3}}
\sqrt{p^2+{\rm sign}(\bar{\delta})(p_xp_y+p_xp_z+p_yp_z)} , ~~~(C) . 
\nonumber
\end{eqnarray}
Thus, we have obtained a linear dispersion laws which demonstrate an essential
anisotropic behavior in the vicinity of the cross-point energy.

\section{CONCLUDING REMARKS}
In this paper, we have developed the generalized $\bf kp$-approach for the
description of the short-range defects with reduced symmetry. We have
considered the peculiarities of the density of states due to the resonant
state contributions as well as those of the alloy bandstructure. In contrast
to the previous considerations, here we have used {\it the matrix form} of 
the potential energy in ${\bf kp}$-equations which leads to the additional
interband mixing. In spite of the simplified model including a few 
phenomenological parameters, we have demonstrated an essential modification 
of the results due to the non-diagonal part of matrix potential.

Next, we discuss the main assumptions used. The above approach is based
on the single-particle description of the electron states which is valid
for many kinds of defects in $A_3B_5$ semiconductors. Although the
general equations (3)-(12) are based on the presumption of low energy of
electrons only, the concrete results are restricted due to the specific 
microscopic potentials used and due to the narrow-gap approximation based 
on the isotropic $6\times 6$ Kane model. The last approximation supposes 
that far-band contributions do not change the results qualitatively. We have 
also used the low-concentration approximation, supposing that the electronic 
states at different impurities do not overlap in Sec. IIB, and we have used 
the Born approximation in Sec. IIC. In order to include an often-discussed 
effect of resonant states on an alloy bandstructure
\cite{7}, one needs to 
use a more complicate self-energy function, e.g. employing 
the coherent 
potential approximation, see Ref.9.
Last but not least, we note 
that the potential $\delta U_{\bf r}$ in Eq.(2) is assumed to be 
spin-independent, i.e. we have neglected the spin-flip processes induced by 
the short-range defect. All these assumptions restrict the quantitative 
description of the phenomena resulting from short-range defects therefor
the results obtained do not related to any concrete case. What this 
paper does, however, it demonstrates clearly that the matrix character of 
the short-range potential energy in Eq.(3) may change the results essentially.

Finally, we suppose that the consideration performed will stimulate
a reexamination the later obtained results both for defects with reduced
symmetry in $A_3B_5$ semiconductors (see discussion in \cite{13}) and for 
bandstructure of $A_3B_5$-based alloys. Application of such kind of approach 
for another materials (e.g. $A_4B_4$-based alloys, see \cite{14}) requires 
a special consideration. In spite of the fact that the obtained results 
demonstrate some peculiarities of $A_3B_5$-based alloys, a more
detail comparison is restricted due to lack of experimental data for the
narrow-gap case under consideration (see recent papers \cite{15}). Since the
components of the matrix potential can not be fixed in the framework
of the {\bf kp}-approach developed, a comparison of such type results with
numerical calculations is also of interest.

\section*{Acknowledgements}
This work was supported by Science Foundation Ireland. I would like to
thank E.P. O'Reilly for useful discussion.

\bigskip $^{*}$ E-mail: {\rm ftvasko@yahoo.com}.

On leave from: Institute of Semiconductor Physics, Kiev, NAS
of Ukraine, 252650, Ukraine


\begin{references}
\bibitem{1}
J. Callaway, {\it Quantum Theory of the Solid State} (Academic, New York,
1967).
\bibitem{2}
G. Lukovsky, Solid State Comm. {\bf 3}, 299 (1965).
\bibitem{3}
V.I. Perel and I.N. Yassievich, Sov. Phys. - JETP {\bf 55}, 143 (1982)
[Zh. Eksp. Teor. Fiz. {\bf 87}, 237 (1982)].
\bibitem{4}
E.V. Bakhanova and F.T. Vasko, Fiz. Tv. Tela {\bf 32}, 86 (1990) [Sov. Phys.
- Solid State {\bf 32}, 47 (1990)].
\bibitem{5}
N.M. Kolchanova, I.D. Loginova, and I.N. Yassievich, Fiz. Tv. Tela {\bf 25},
1650 (1983) [Sov. Phys. - Solid State {\bf 25}, 952 (1983)].
\bibitem{6}
A.V. Germanenko and G.M. Minkov, Physica Status Solidi (b), {\bf 184}, 9
(1994).
\bibitem{7}
 Semicond. Sci. and Technol. {\bf 17}, N8 (2002) Special Issue:
III-N-V Semiconductor Alloys.
\bibitem{8}
It should be noted that the case under consideration is a multi-band
problem because a strongly singular potential energy, which is proportional 
not only to $\delta ({\bf r})$ but also to $\nabla\delta ({\bf r})$, appears 
under the
standard effective mass transformation of Eq.(3) according to Ref.1.
\bibitem{9}
E.N. Economou, {\it Green's Functions in Quantum Physics} (Springer-Verlag,
Berlin, 1983).
\bibitem{10}
R.J. Elliott, J.A. Krumhansl, and P.J. Leath, Rev. Mod. Phys. {\bf 46},
465 (1974).
\bibitem{11}
D.C. Hutchings and B.S. Wherrett, Phys. Rev. B {\bf 48}, 2418 (1994).
\bibitem{12}
Taking into account the spin-split-off band or considering a defect which is
non-symmetric along a diagonal axis ($[111]$ or $[110]$ directions)
one can obtain a bowing of extreme energies.
\bibitem{13}
Y. Zhang and W. Ge, Journ. of Luminescence, {\bf 85}, 247 (2000).
\bibitem{14}
Y.I. Ravich and S.A. Nemov, Semiconductors, {\bf 36}, 1 (2002);
B.A. Akimov, A.V. Dmitriev, D.R. Khokhlov, and L.I. Ryabova, Physica Status
Solidi (a), {\bf 137}, 9 (1993)
\bibitem{15}
B.N. Murdin, A.R. Adams, P. Murzyn, C.R. Pidgeon, I.V. Bradley, J.P.R. Wells,
Y.H. Matsuda, N. Miura, T. Burke, and A.D. Johnson, Appl. Phys. Lett.
{\bf 81}, 256 (2002); B.N. Murdin, M. Kamal-Saadi, A. Lindsay, E.P. O'Reilly,
A.R. Adams, G.J. Nott, J.G. Crowder, C.R. Pidgeon, I.V. Bradley, J.P.R. Wells,
T. Burke T, A.D. Johnson, and T. Ashley, Appl. Phys. Lett. {\bf 78}, 1568
(2001).
\end{references}
\end{document}